# A Comparison of Electronic, Dielectric, and Thermoelectric Properties of Monolayer of HfX$_2$N$_4$(X = Si, Ge) through First-Principles Calculations

Chayan Das [a], Abhishek [b], Dibyajyoti Saikia [a], Appala Naidu Gandi [b], Satyajit Sahu [a]*

[a] *Department of Physics, Indian Institute of Technology Jodhpur, Jodhpur 342037, India*

[b] *Department of Metallurgical and Materials Engineering, Indian Institute of Technology Jodhpur, Jodhpur 342037, India*

## Abstract

The newly emerged two-dimensional (2D) materials family of MSi$_2$N$_4$, where M is a transition metal atom (i.e., Mo, W, etc.), has the potential to be named after the conventional and very popular transition metal di-chalcogenides (TMDC), which got their reputation for having bandgap tunability and high mobility. The HfSi$_2$N$_4$ and HfGe$_2$N$_4$ 2D materials are members of the MSi$_2$N$_4$ family and possess very good figure of merit (*ZT*) and have high mobility, proving their suitability for thermoelectric applications. The HfSi$_2$N$_4$ and HfGe$_2$N$_4$ showed considerable *ZT* of 0.90 and 0.89, respectively, for p-type and 0.83 and 0.79 for n-type, at 900 K along with high mobility according to the solutions obtained after solving the Boltzmann Transport Equation (BTE). The HfGe$_2$N$_4$ also showed a *ZT* of 0.84 at 600 K and 0.68 at 300 K, which is also excellent for low-temperature operation. The bandgaps (BG) obtained for HfSi$_2$N$_4$ and HfGe$_2$N$_4$ according to the Heyd-Scuseria-Ernzerhof (HSE) approximation were 2.89 eV and 2.75 eV. The first absorption peak showed in the blue region of the visible spectrum; from this, their usefulness in visible range photodetectors can also be inferred.

# 1. Introduction

The extraordinary electrical [1], thermal [2], and optical [2] properties exhibited by two-dimensional (2D) materials have been witnessed in recent times, which attracted considerable interest in exploring more of these materials [3–6]. $MoSi_2N_4$ monolayer (ML), a member of the $MX_2N_4$ family, was recently synthesized using the CVD method on a centimeter scale [7] with high strength and stability [8], which motivates researchers to explore more in this direction. Thermoelectrics can be used for the efficient utilization of energy resources. As non-renewable energy sources are depleting, people need to think more and more about efficient utilization of existing resources. In this work, Boltzmann transport theory along with the first-principles calculations were considered to evaluate the 2D MLs of $HfSi_2N_4$ and $HfGe_2N_4$. We obtained considerable Seebeck coefficients ($S$) and $ZT$. Here, $ZT$ is the figure of merit, while the Seebeck coefficient quantifies the voltage produced when a gradient of temperature is used at the two extreme points of the material. An ideal thermoelectric material should have a combination of high electrical conductivity ($\sigma$) and $S$, as well as a low thermal conductivity ($k = k_e + k_l$). The overall thermal conductivity is affected by its two components, i.e., originating from electrons ($k_e$) and originating from (phonons) lattice ($k_l$). The 2D MLs exhibit exceptional stability, superior mobility, and favorable optoelectronic characteristics [8]. The crystal structure can be viewed as a metal atom (M) sandwiched between two nitrogen (N) atoms, and the whole section was again sandwiched between two buckled honeycomb XN layers where X = Si or Ge. Here M is the transition metal atom (Mo, W, Cr, Hf, etc.). One possible method for their synthesis is chemical vapor deposition (CVD). Recently, $MoSi_2N_4$ was synthesized by Yi-Lun and his coworkers by using CVD [7]. The mobility values of $MoSi_2N_4$ were predicted to be 490 $cm^2V^{-1}s^{-1}$ and 2190 $cm^2V^{-1}s^{-1}$ for n and p-type and for $WSi_2N_4$ the values were 668 $cm^2V^{-1}s^{-1}$ and 1385 $cm^2V^{-1}s^{-1}$ for n and p-type semiconductors [8,9]. Considerable investigation has also been conducted on two-dimensional transition metal chalcogenide materials. The $MoSe_2$ ML was reported with an exceptionally low $ZT$ of 0.2 and a high $k_l$ of 60 $Wm^{-1}K^{-1}$. Qing Tan and his coworkers experimentally reported a low $k_l$ of 1.25 $Wm^{-1}K^{-1}$ with SnS ML; however, the $ZT$ value was found to be about 0.6 at 900 K [10]. $TiS_2$ ML was reported with a $k_l$ of more than 4 $Wm^{-1}K^{-1}$ with $ZT$ value of 0.5 [11]. $WS_2$ ML showed a very high $k_l$ of 23.15 $Wm^{-1}K^{-1}$ along with a $ZT$ of 0.66 at 900 K [12]. $HfS_2$ ML showed a low $k_l$ of 2.83 $Wm^{-1}K^{-1}$ with a considerable $ZT$ of 0.90 [13]. ZrOS Janus ML was also reported with a $ZT$ of 0.82 with a very low $k_l$ of 1.5 $Wm^{-1}K^{-1}$ [14]. Apart from the 2D transition metal chalcogenides, $MoSi_2N_4$, and $MoGe_2N_4$, were reported with a ZT of 0.42 and 0.94 along with $k_l$ of 260 Wm$^{-1}$K$^{-1}$ and 80 Wm$^{-1}$K$^{-1}$ [15]. WSi$_2$N$_4$ and WGe$_2$N$_4$ were also reported with a ZT of 0.72 and 0.91, along with a $k_l$ of 114.9 Wm$^{-1}$K$^{-1}$ and 9.41 Wm$^{-1}$K$^{-1}$ [9]. Research is still going on with the new 2D materials to improve their performance as thermoelectric materials. Here, we studied the electrical, optical, and thermoelectric properties of HfSi$_2$N$_4$ and HfGe$_2$N$_4$ using the first-principles calculations associated with DFT and BTE. The HfGe$_2$N$_4$ ML exhibited an excellent ZT of 0.90 (0.83 for n-type), which is better in comparison to HfSi$_2$N$_4$, which exhibited a ZT of 0.89 (0.79 for n-type) at 900 K for p-type doping. So, a strategic comparison of thermoelectric and optical properties is needed to understand the physical and chemical properties, which play an important role in the difference in their properties. The HfSi$_2$N$_4$ and HfGe$_2$N$_4$ possess similar bandgaps. As a result, they show absorption peaks in the same wavelength range, which was blue in the visible range, i.e. 490 nm and 493 nm, respectively.

## 2. Methodology

First-principles calculations were performed with the projector augmented wave (PAW) [16,17] method implemented in Vienna ab-initio Simulation Package (VASP) [18] with exchange-correlation interaction of Perdew-Burke-Ernzerhof (PBE) functionals [19]. The DFT-D3 method of Grimme was used to include the van der Waals (vdW) interactions. A vacuum of 30 Å thickness was maintained along the z-direction to avoid the interaction of two layers. The geometry optimization was performed using a gamma (Γ) centered 12×12×1 k-meshes. The energy cutoff was set to 520 eV, and the electronic and ionic convergence criteria were set to 10$^{-6}$ eV and 10$^{-5}$ eV/ Å. The phonon band structure was obtained from the Phonopy package in combination with VASP using 2×2×1 supercells with forces estimated employing Γ centered 9×9×1 k-meshes. More accurate representation of the electronic band structure is obtained employing the Heyd-Scuseria-Ernzerhof (HSE) hybrid functionals [20]. 15×15×1 k-meshes were utilized, along with the previously determined ionic, electronic convergence and kinetic energy cutoffs levels. The ab initio molecular dynamics (AIMD) calculation, mobility, and relaxation time calculations were performed using the Quantum Espresso (QE) Package [21,22]. In these calculations, 15×15×1 k-meshes were used for the Brillouin zone integrations and an energy cutoff value of 50 Ry was used for the electronic wave functions. 10$^{-9}$ Ry was used as electronic energy convergence criterion in these calculations. The optical response as a function of frequency was evaluated using the Bethe-Salpeter equation (BSE) [8]. In these hybrid BSE (HSE+BSE) calculations the 9×9×1 Γ centred k-meshes were used.

The many-body perturbation theory calculation necessitates a significant number of unoccupied states. As a result, 128 electronic bands were taken into account. The refractive index (η) and extinction coefficient ($K$) were determined using the following formulae.

$$\eta = \left[\frac{\left\{(\varepsilon_r^2 + \varepsilon_i^2)^{1/2} + \varepsilon_r\right\}}{2}\right]^{1/2} \quad (1)$$

$$K = \left[\frac{\left\{(\varepsilon_r^2 + \varepsilon_i^2)^{1/2} - \varepsilon_r\right\}}{2}\right]^{1/2} \quad (2)$$

And using them, the absorption coefficient (α) was also determined.

$$\alpha = \frac{2K\omega}{C} \quad (3)$$

Here, $\varepsilon_r$, $\varepsilon_i$ are real and imaginary parts of the dielectric function, and $\omega$, and $C$ are frequency, and speed of light, respectively. Thermoelectric transport coefficients i.e. Seebeck coefficient ($S_{l,m}$), electrical conductivity ($\sigma_{l,m}$), and electronic thermal conductivity ($k_{l,m}$) were obtained from the semi-classical BTE with the constant scattering time approximation (CRTA) implemented in BoltzTraP code [23].

$$\sigma_{l,m} = \frac{1}{\Omega}\int \sigma_{l,m}(\varepsilon)\left[-\frac{\Delta f_\phi(T,\varepsilon)}{\Delta\varepsilon}\right]d\varepsilon \quad (4)$$

$$k_{l,m}(T,\phi) = \frac{1}{e^2 T\Omega}\int \sigma_{l,m}(\varepsilon)(\varepsilon-\phi)^2\left[-\frac{\Delta f_\phi(T,\varepsilon)}{\Delta\varepsilon}\right]d\varepsilon \quad (5)$$

$$S_{l,m}(T,\phi) = \frac{(\sigma^{-1})_{n,l}}{eT\Omega}\int \sigma_{n,m}(\varepsilon)(\varepsilon-\phi)\left[-\frac{\Delta f_\phi(T,\varepsilon)}{\Delta\varepsilon}\right]d\varepsilon \quad (6)$$

Here, $e, \phi, \Omega, T$ are the electron charge, chemical potential, unit cell volume, and temperature, respectively. Lattice contribution to the thermal conductivity ($k_l$) is calculated using the phono3py package in combination with quantum espresso by evaluating the third-order force constants with the finite displacement (0.06 Å) method. In phono3py, a 2×2×1 supercell was used, and the forces were estimated employing 9×9×1 k-meshes.

## 3. Result and Discussion

### 3.1. Structure and Stability

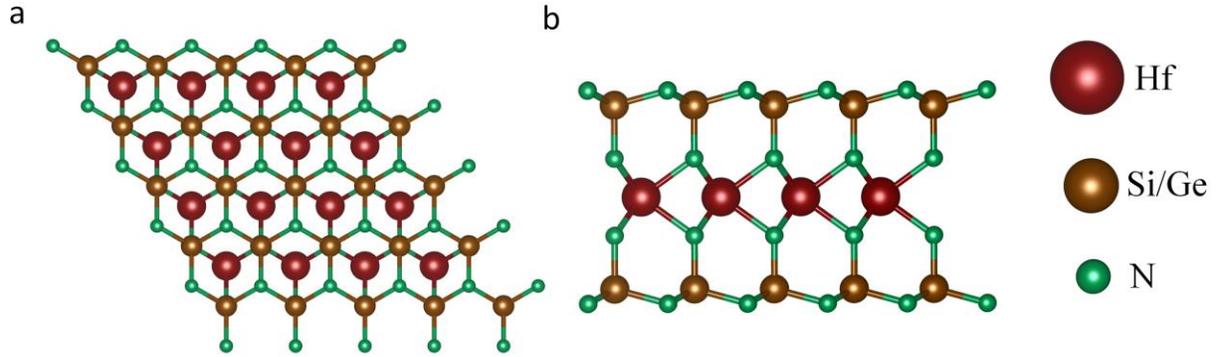

**Figure 1:** Visualization of (a) top and (b) side view of HfX$_2$N$_4$ (X = Si, Ge) ML with 4×4×1 supercell.

The HfX$_2$N$_4$ ML, where X can be Si or Ge, can be seen as a ML of HfN$_2$ positioned between two honeycomb layers made of SiN or GeN. The three layers are arranged in a stacked formation. The Hf atom is located in the center of a trigonal prism structure composed of six Si atoms. The HfN$_2$ layer is linked to the (SiN/GeN) layers via vertically aligned Si-N connections (**Figure 1(a)**). HfX$_2$N$_4$ possesses a hexagonal structure (**Figure 1(b)**), which belongs to the space groups of P-6m2 (No. 187) [8]. The optimized lattice constant values for HfSi$_2$N$_4$ were consistent with prior findings [8]. The lattice constants for HfGe$_2$N$_4$ were also optimized. **Table 1** represents the summary of optimized lattice constants ($a$), thickness (t), bond lengths (d), and angles ($\theta$) for the two MLs. It was observed that the bond length in the HfGe$_2$N$_4$ is greater compared to that of HfSi$_2$N$_4$ as the atomic radius of Ge is greater than the atomic radius of Si.

**Table 1:** Calculated lattice constants ($a$), thickness (t), bond lengths (d), and angles ($\theta$) for 2D ML of HfSi$_2$N$_4$ and HfGe$_2$N$_4$.

| Monolayer | $a(=b)$ (Å) | Thickness (Å) | d(Hf-N) (Å) | d(X-N) (Å) | θ(Hf-N-Hf) | θ(N-X-N) | θ(Hf-N-X) |
|---|---|---|---|---|---|---|---|
| HfSi$_2$N$_4$ | 3.01 | 11.36 | 2.14 | 1.75 | 71.57 | 105.58 | 125.78 |
| HfGe$_2$N$_4$ | 3.14 | 11.50 | 2.18 | 1.90 | 67.33 | 108.86 | 123.66 |

Cohesive energy provides insight into the structural stability of a material. The cohesive energy ($E_{co}$) was determined using the provided formula: $E_{co} = \{(E_{Hf} + 2 \times E_X + 4 \times E_N) - E_{HfX_2N_4}\}/7$. The $E_{HfX_2N_4}$ is the energy of HfX$_2$N$_4$ ML and, $E_{Hf}$, $E_X$, and $E_N$ are the atomic energy of single Hf, X, and N, respectively. The obtained $E_{co}$ per atom was evaluated as 1.11 eV and 0.50 eV for HfSi$_2$N$_4$ and HfGe$_2$N$_4$ MLs indicating their thermodynamic stability. We assessed the dynamical stability of HfSi$_2$N$_4$ and HfGe$_2$N$_4$ by calculating their phonon dispersion curves along the Γ-M-K-Γ path and shown in **Figure 2**. No imaginary frequencies in the phonon band structures reaffirms the thermodynamic stability of these structures. The maximum phonon frequency found in HfSi$_2$N$_4$ and HfGe$_2$N$_4$ were 29.89 THz and 26.00 THz, respectively, suggesting that the HfGe$_2$N$_4$ has a lower maximum frequency compared to HfSi$_2$N$_4$. Twenty-one different vibrational modes were found as the ML's unit cells contain seven atoms. The first three modes from the bottom are called the acoustic modes, while the following eighteen modes are optical modes. The acoustic vibrational modes are in-plane longitudinal acoustic (LA) mode, transverse acoustic (TA) mode, and out-of-plane mode (ZA) from the bottom. The optical bands in the HfGe$_2$N$_4$ ML were found to overlap with the acoustic modes, in contrast to the HfSi$_2$N$_4$ ML, where the acoustic bands are separate. From the phonon density of states (PhDOS), it was found that the Hf atom contributed more towards acoustic modes compared to optical modes, and Si/Ge and N contributed more to the optical modes compared to acoustic modes. To ensure the practical applicability of a thermoelectric device operating at a high temperature of 900 K, we performed the ab initio molecular dynamics (MD) simulation for more than 4500 femtoseconds for both HfSi$_2$N$_4$ and HfGe$_2$N$_4$ under a micro-canonical ensemble at a temperature of 900 K. **Figures 2c and 2d** show the thermal stability of these two systems at 900 K.

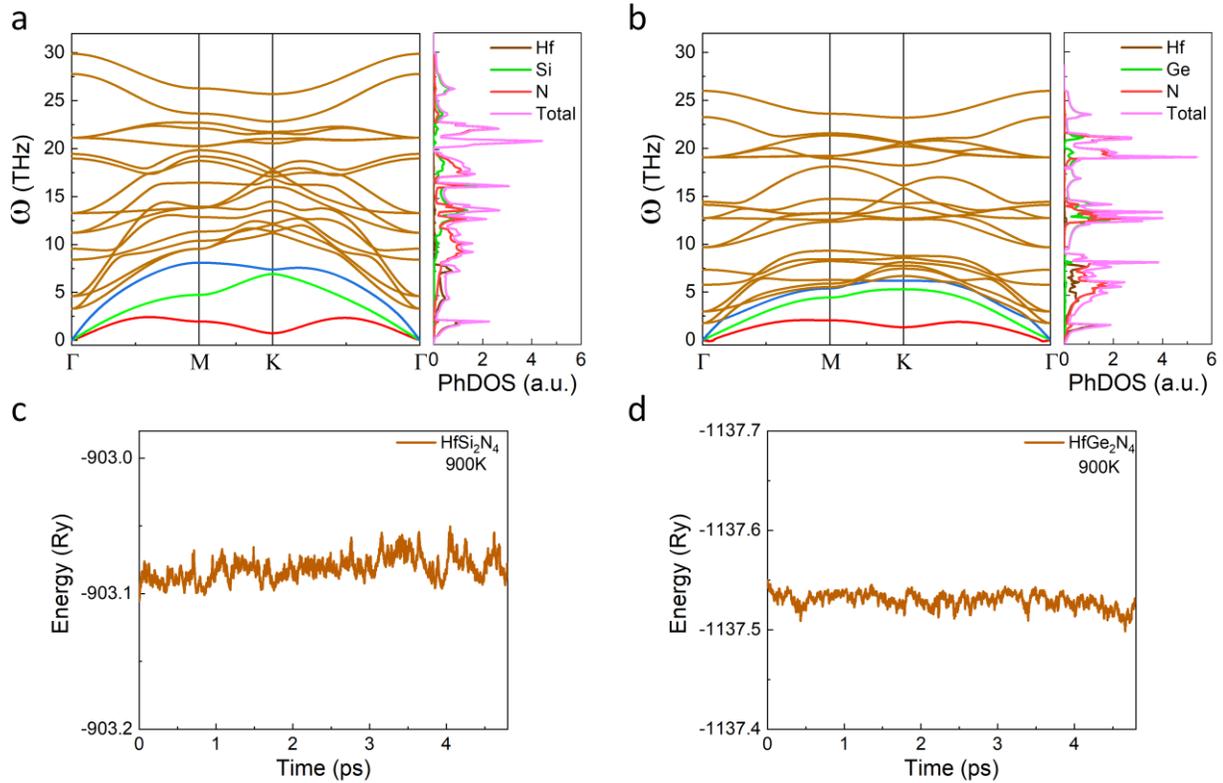

**Figure 2:** Calculated phonon band structure and phonon density of states in a) $HfSi_2N_4$ and b) $HfGe_2N_4$ ML. The red, green, and blue colored bands represent the ZA, TA, and LA modes, respectively. Energy fluctuation at 900 K estimated in MD simulations are shown in c) $HfSi_2N_4$ and d) $HfGe_2N_4$.

### 3.2. Electronic Properties

**Figures 3a and 3b** show the electronic band structure from -4 eV to 4 eV of energy range. The calculation was performed along the Γ-M-K-Γ path for the $HfSi_2N_4$ and $HfGe_2N_4$ ML. The obtained BGs are 1.79 eV and 1.70 eV, respectively, for $HfSi_2N_4$ and $HfGe_2N_4$ according to PBE approximation. **Figures 3c and 3d** show the more accurate band structure of $HfSi_2N_4$ and $HfGe_2N_4$ obtained using hybrid functionals (HSE06), and the corresponding BG values are 2.89 eV and 2.75 eV, respectively. The BG values achieved for both materials match well with the values suggested by calculations conducted by Yadav et al. [24]. The nature of both the band structures is indirect, and both the MLs show the conduction band minima (CBM) between Γ and M point. The valance band maximum (VBM) is between the M and K points in the case of $HfSi_2N_4$. However, in the case of

HfGe$_2$N$_4$, the VBM is situated between the K and Γ points. From the DOS and local density of states (LDOS) obtained for the different elements, it was observed that for both the HfSi$_2$N$_4$ and HfGe$_2$N$_4$ MLs, the VBM is primarily contributed by the *p* orbitals of N atoms and *d* orbitals of Hf atoms. But CBM is primarily contributed by *d* of Hf atoms. It has been noted that in both the MLs, the element N has a more significant impact on the valence band, while Hf has a greater effect on the conduction band. The *p* orbital of Silicon (Si) and Germanium (Ge) make similar contributions to both the valance bands and conduction bands. Improved DOS calculations utilizing hybrid functionals (HSE06) exhibited a similar trend as those estimated using PBE functionals.

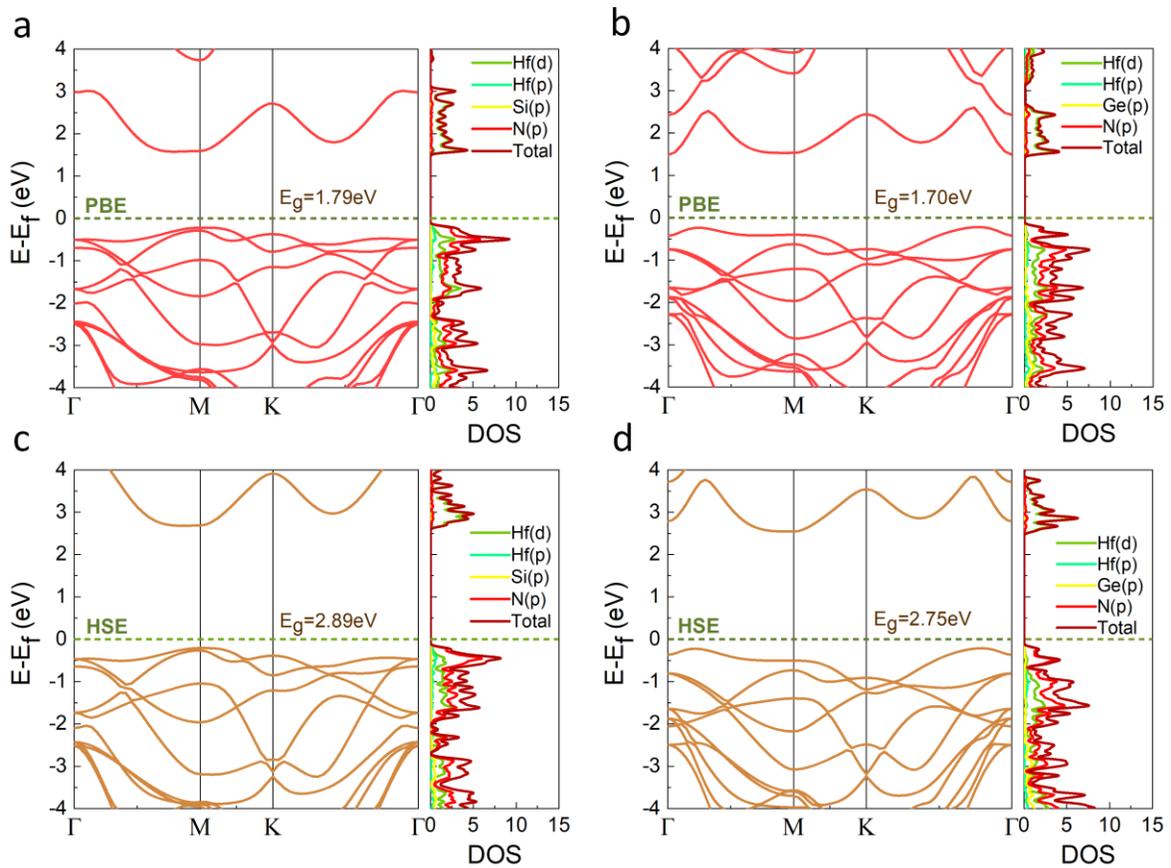

**Figure 3:** Calculated band structure and local density of states of the a) HfSi$_2$N$_4$ and b) HfGe$_2$N$_4$ MLs, obtained from PBE functionals, and c) HfSi$_2$N$_4$ and d) HfGe$_2$N$_4$ obtained from the HSE06 functionals.

### 3.3. Mobility and relaxation time

Bardeen and Shockley's deformation potential theory was used to calculate the mobility of electrons and holes according to PBE with QE. Here, the band shift was considered with changing strain, which was stated as the deformation potential of individual crystals [25]. The individual mobilities of individual 2D systems can be determined by calculating the deformation potentials and taking into account the effective mass theorem, which is expressed as

$$\mu_{2D} = \frac{e\hbar^3 C_{2D}}{K_B T m_e^* m_a (E_l^2)}. \qquad (7)$$

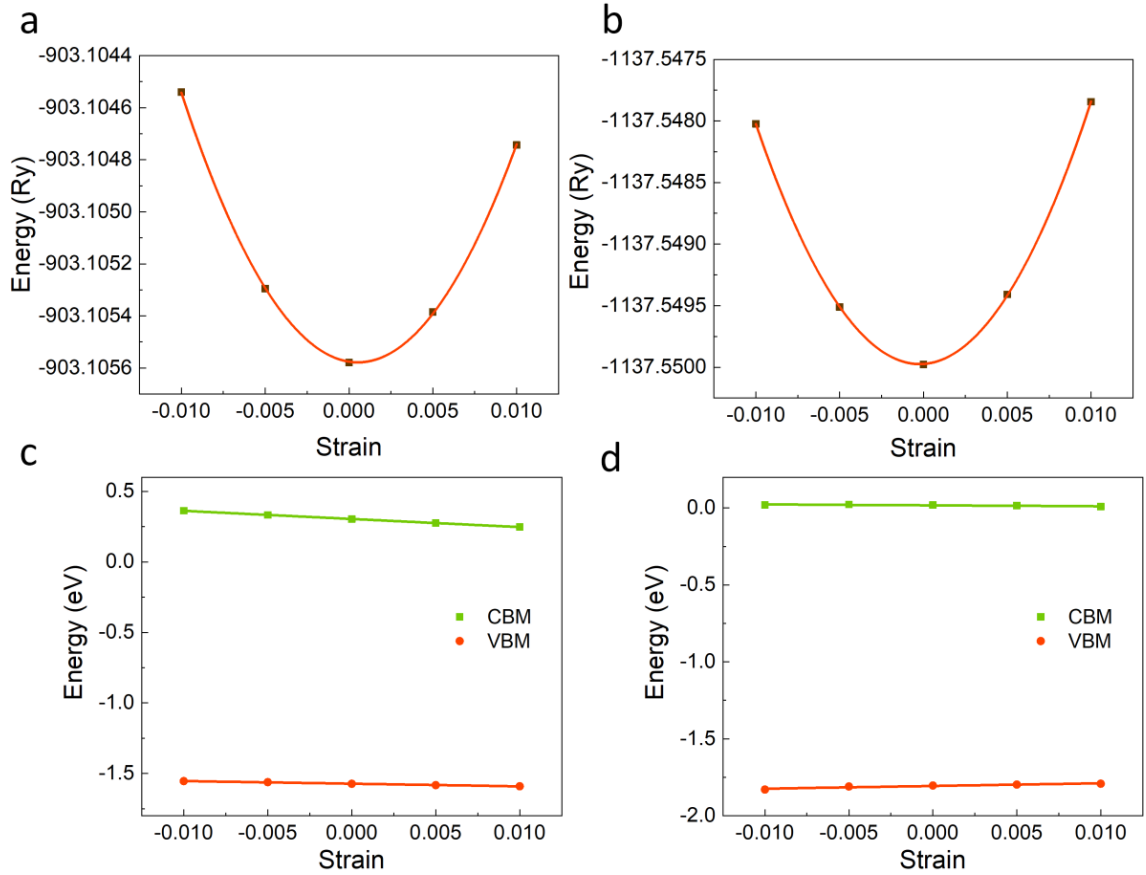

**Figure 4:** Variation of total energy with respect to uniaxial strain in a) HfSi$_2$N$_4$ and b) HfGe$_2$N$_4$. To determine the deformation potential, the VBM and CBM shift in c) HfSi$_2$N$_4$ and d) HfGe$_2$N$_4$ with respect to uniaxial strain has been calculated.

The elastic constant of 2D MLs is represented by $C_{2D}$ of HfSi$_2$N$_4$ and HfGe$_2$N$_4$. We can derive this from the following formula $C_{2D} = (\frac{\partial^2 E}{\partial \varepsilon^2}/A_0)$. We plot the energy values for different uniaxial strains and fit a parabola to determine $C_{2D}$ from the plot of HfSi$_2$N$_4$ (**Figure 4a**) and for HfGe$_2$N$_4$

(**Figure 4b**). Here $A_0$ represents the area of the individual MLs in the absence of strain, T denotes the temperature in Kelvin, $K_B$ represents the Boltzmann constant, and $m^*$ signifies the effective mass, which is estimated using $m^*(=\frac{1}{\hbar^2}\frac{d^2E}{dk^2})$ for electron and holes, and $m_a = \sqrt{m_x^* m_y^*}$ is the geometric average of in-plane effective masses. As the effective mass also possesses a second-order derivative, we can get it from the band edge of VBM (for hole) and CBM (for electron) by fitting a parabola. The obtained $C_{2D}$ is 388.08 Nm⁻¹, and hole mobility ($\mu_h$) is 582.35 cm²V⁻¹s⁻¹ for HfSi₂N₄. The $C_{2D}$ and $\mu_h$ of HfGe₂N₄ were calculated to be 330.71 Nm⁻¹ and 1870.08 cm²V⁻¹s⁻¹, and are in agreement with the previous results [8,26]. The deformation potential ($E_I = \frac{\partial V}{\partial \varepsilon}$), was calculated from the change in band edges (V) with the application of strain ($\varepsilon$). **Figures 4(c) and (d)** represent the change of CBM and VBM band edges for HfSi₂N₄ and HfGe₂N₄. Relaxation time ($\tau$) for electrons and holes for HfSi₂N₄ and HfGe₂N₄ can be evaluated from the following expression: $\tau = \frac{\mu m^*}{e}$. The complete set of parameters acquired for both structures is presented in **Table 2**, and $m_0$ represents the rest mass of the electron. The materials of our interest show very high mobility ($\mu$), which exceeds the mobility of many reported 2D materials by more than ten times, i.e., MoS₂ (72 cm²V⁻¹s⁻¹) and WS₂ (21 cm²V⁻¹s⁻¹) [27]. The $\mu$ of these materials is similar to that of boron phosphorous (BP) and boron arsenide (BAs); however, they are unsuitable for thermoelectric applications due to their exceedingly high $\kappa_l$ of approximately 4×10² W/mK [28].

**Table 2:** Calculated $C_{2D}$, $E_I$, $m^*$, $\mu$, and $\tau$ of ML of HfSi₂N₄ and HfGe₂N₄ for both electrons and holes according to PBE.

| Crystal | $C_{2D}$ (N/m) | $E_{I,(electron)}$ (eV) | $E_{I,(hole)}$ (eV) | $m_e^*$ /$m_0$ | $m_h^*$ /$m_0$ | $\mu_e$ (cm²V⁻¹s⁻¹) | $\mu_h$ (cm²V⁻¹s⁻¹) | $\tau_e$ (s) × $10^{-13}$ | $\tau_h$ (s) × $10^{-13}$ |
|---|---|---|---|---|---|---|---|---|---|
| HfSi₂N₄ | 388.08 | 5.78 | 1.87 | 1.16 | 2.017 | 181.20 | 582.35 | 1.20 | 6.68 |
| HfGe₂N₄ | 330.71 | 4.39 | 1.45 | 1.30 | 1.341 | 216.78 | 1870.08 | 1.61 | 14.26 |

### 3.4. Dielectric properties

The response of dielectric functions was evaluated using the Bethe-Salpeter equation (BSE) [29] in accordance with the HSE calculations. The variation of the imaginary dielectric function is shown in **Figure 5a**, and the real part is shown in **Figure 5b** with respect to the energy. At 2.52 eV for HfSi$_2$N$_4$ and 2.49 eV for HfGe$_2$N$_4$, the $\varepsilon_i$ showed their first peak. The phenomena are deducible from the band structures that appear, as both the MLs HfSi$_2$N$_4$ and HfGe$_2$N$_4$ possess comparable bandgaps having only a difference of 0.14 eV. The second peaks with high magnitude were situated at 2.94 eV for HfSi$_2$N$_4$ and 3.03 eV for HfGe$_2$N$_4$. At 2.42 eV and 2.38 eV, the $\varepsilon_r$ showed the first peaks for HfSi$_2$N$_4$ and HfGe$_2$N$_4$. The absorption spectra exhibited distinct peaks at 2.53 eV and 2.51 eV, corresponding to $\alpha$ values of $1.01 \times 10^5$ cm$^{-1}$ and $0.53 \times 10^5$ cm$^{-1}$ for HfSi$_2$N$_4$ and HfGe$_2$N$_4$, respectively (**Figure 5c**), and they match with the already reported values in the literature [14,30]. Secondary peaks with higher magnitude were also found at 3.04 eV for HfSi$_2$N$_4$ and 2.96 eV for HfGe$_2$N$_4$. The dependence of η with energy is shown in **Figure 5d**, the obtained value of η is about 1.55 and 1.64 for HfSi$_2$N$_4$ and HfGe$_2$N$_4$ at zero energy, respectively. The first peaks for both HfSi$_2$N$_4$ and HfGe$_2$N$_4$ are located at energy levels of 2.43 eV and 2.39 eV, whereas the secondary peaks are observed at higher energy levels of 2.98 eV and 2.85 eV for HfSi$_2$N$_4$ and HfGe$_2$N$_4$. After 10 eV, the η of HfSi$_2$N$_4$ and HfGe$_2$N$_4$ become nearly identical.

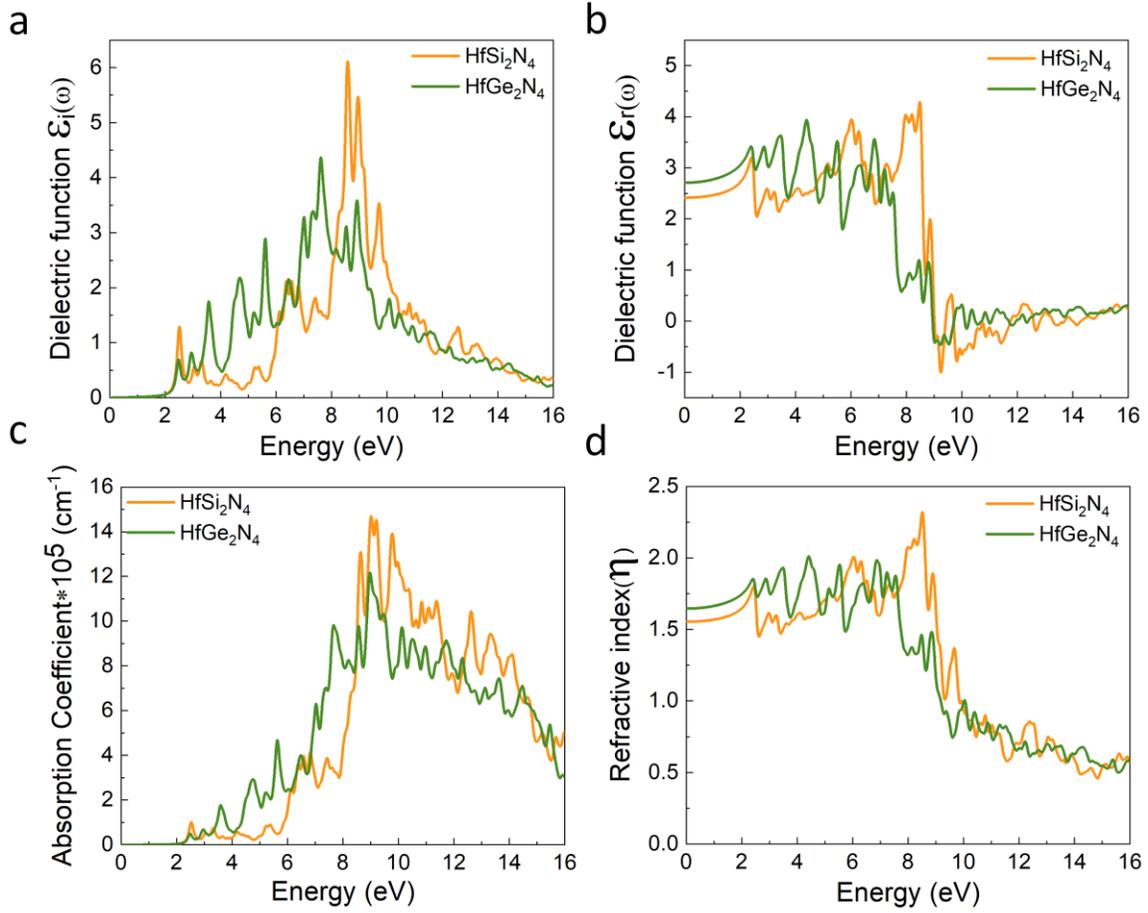

**Figure 5:** Plot of a) imaginary ($\varepsilon_i$), and b) real ($\varepsilon_r$) component of the dielectric function, c) absorption coefficient ($\alpha$), and d) refractive index ($\eta$) as a function of energy based on the HSE calculations.

## 3.5. Thermoelectric Properties

**Figure 6a** shows the variation of the Seebeck coefficient ($S$) of the HfSi$_2$N$_4$ ML as a function of chemical potential ($\phi$) at three specific temperatures: 300 K, 600 K, and 900 K. The highest Seebeck coefficient ($S$) found for HfSi$_2$N$_4$ monolayer (ML) is -2924.99 µVK$^{-1}$ for n-type carriers ($\phi > 0$) and 2994.27 µVK$^{-1}$ for p-type carriers ($\phi < 0$) at a temperature of 300 K. The absolute value of $S$ reached its peak at 300 K and subsequently decreases as temperature increases. **Figure 6b** displays the relationship between the relaxation time-scaled electrical conductivity ($\sigma/\tau$) with respect to $\phi$. No significant change was observed in the value of $\sigma/\tau$ with respect to the change in temperature. We observed that the $\sigma/\tau$ for the p-type is lower than that of the n-type in ML

HfSi$_2$N$_4$. **Figure 6c** illustrates the relationship between the relaxation time-scaled power factor (PF = $S^2\sigma/\tau$) and ϕ for the HfSi$_2$N$_4$ ML. The maximum power factor achieved in a monolayer of HfSi$_2$N$_4$ was 38.43×10$^{10}$ Wm$^{-2}$K$^{-1}$s$^{-1}$ for p-type carriers (37.88×10$^{10}$ Wm$^{-2}$K$^{-1}$s$^{-1}$ for p-type) at a temperature of 900 K. **Figures 6d, e, and f** illustrate the changes in $S$, $\sigma/\tau$, and $S^2\sigma/\tau$ for the HfGe$_2$N$_4$ ML as a function of ϕ. The $S$ of the HfSi$_2$N$_4$ ML decreases gradually as the temperature increases. The maximum achieved power factor (PF) for the monolayer (ML) of HfGe$_2$N$_4$ is 36.87 ×10$^{10}$ Wm$^{-2}$K$^{-1}$s$^{-1}$, specifically for the n-type carriers. The power factor obtained is 13.02 ×10$^{10}$ Wm$^{-2}$K$^{-1}$s$^{-1}$ at a temperature of 900 K for the p-type carriers. The efficiency of n-type doping in HfGe$_2$N$_4$ for thermoelectric applications is significantly higher. The Seebeck coefficient ($S$) of HfSi$_2$N$_4$ ML was found to be greater than that of HfGe$_2$N$_4$, which can be attributed to the higher bandgap (BG) value of HfSi$_2$N$_4$ compared to HfGe$_2$N$_4$.

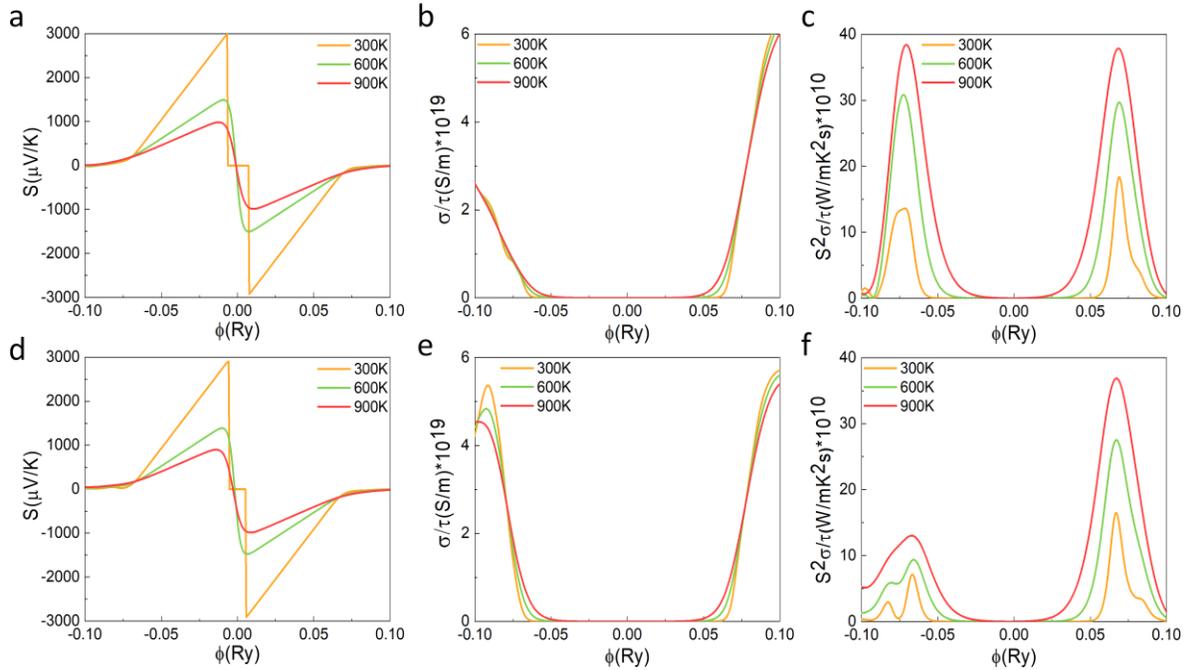

**Figure 6:** Variation of (a) $S$ (b) $\sigma/\tau$, and (c) PF for HfSi$_2$N$_4$ with respect to ϕ, and variation of (d) $S$ (e) $\sigma/\tau$, and (f) PF for HfGe$_2$N$_4$ with respect to chemical potential ϕ.

### 3.6. Lattice thermal conductivity ($\kappa_l$)

**Figures 7a and 7b** represent the relationship between the variation of $\kappa_l$ with temperature for ML of HfSi$_2$N$_4$ and HfGe$_2$N$_4$. The thermal conductivity ($\kappa_l$) of HfGe$_2$N$_4$ was found to be significantly lower (22.41 Wm$^{-1}$K$^{-1}$) at 300 K compared to HfSi$_2$N$_4$ (38.76 Wm$^{-1}$K$^{-1}$), as well as lower than well-known other 2D TMDCs such as MoS$_2$ (34.5 Wm$^{-1}$K$^{-1}$) [31] and WS$_2$ (72 Wm$^{-1}$K$^{-1}$) [12]. Both structures exhibit a drop in $\kappa_l$ as temperature increases. The monolayer (ML) of HfGe$_2$N$_4$ demonstrated significantly reduced $\kappa_l$ compared to the monolayer of WSi$_2$N$_4$, even at elevated temperatures [9].

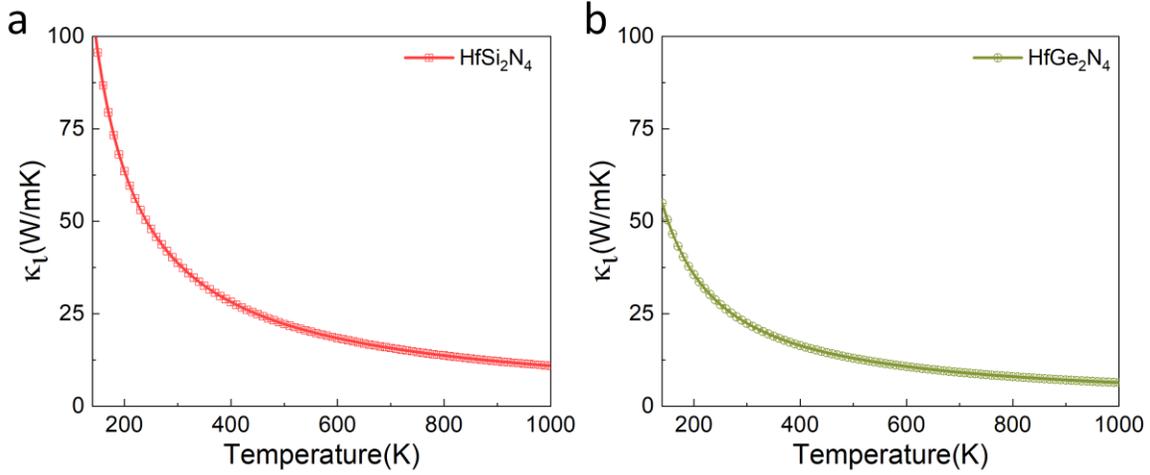

**Figure 7:** Temperature-dependent variation of $\kappa_l$ for ML (a) HfSi$_2$N$_4$ and (b) HfGe$_2$N$_4$.

**Figure 8** shows the variation of group velocity ($G_v$) and Grüneisen parameters ($\gamma$) with respect to frequency for both MLs. The lattice thermal conductivity difference can be found from group velocity analysis. **Figure 8a** shows a maximum $G_v$ of about $11.0 \times 10^3$ m/s for the HfSi$_2$N$_4$ ML precisely because of the longitudinal acoustic mode. Similarly, the HfGe$_2$N$_4$ ML showed a maximum $G_v$ of about $9.1 \times 10^3$ m/s (**Figure 8b**), which is also specifically because of the longitudinal acoustic mode. Since the value of $\kappa_l$ is directly proportional to $G_v$, it is straightforward to explain the difference in values of $\kappa_l$ for the MLs. Due to the significantly greater $G_v$ value of HfSi$_2$N$_4$ ML compared to HfGe$_2$N$_4$, the thermal conductivity ($\kappa_l$) is also much higher in HfSi$_2$N$_4$. It was found that the presence of both acoustic mode and optical mode phonon bands are distinct with no overlap by analyzing the phonon dispersion curve of HfSi$_2$N$_4$. Therefore, the acoustic and optical phonon scattering is less probable to occur, which is evident in $\kappa_l$ values for HfSi$_2$N$_4$ and HfGe$_2$N$_4$ ML. A decrease in the overall phonon frequencies was found due to the presence of the heavier Ge atom in HfGe$_2$N$_4$ when compared to HfSi$_2$N$_4$. In HfGe$_2$N$_4$, there is a simultaneous presence of low-lying optical phonons and acoustic phonons, with some degree of overlap. It has been observed that low-lying optical phonons are close to acoustic phonons and sometimes overlap in HfGe$_2$N$_4$. This leads to enhanced scattering of acoustic phonons by the optical phonons, which is responsible for the lowering value of $\kappa_l$ for the HfGe$_2$N$_4$ ML [29]. The calculation of the Grüneisen parameter ($\gamma$) is also included in our study. Slacks equation states that the $\kappa_l$ varies inversely to the square of $\gamma$ [33], which can be inferred from our analysis. From **Figure 8c and d** we can see that the HfGe$_2$N$_4$ possessed a higher (anharmonicity) value ($\gamma$) compared to HfSi$_2$N$_4$, which is follows the same trend of $\kappa_l$ i.e., HfSi$_2$N$_4$ possesses higher value of $\kappa_l$ compared to HfGe$_2$N$_4$

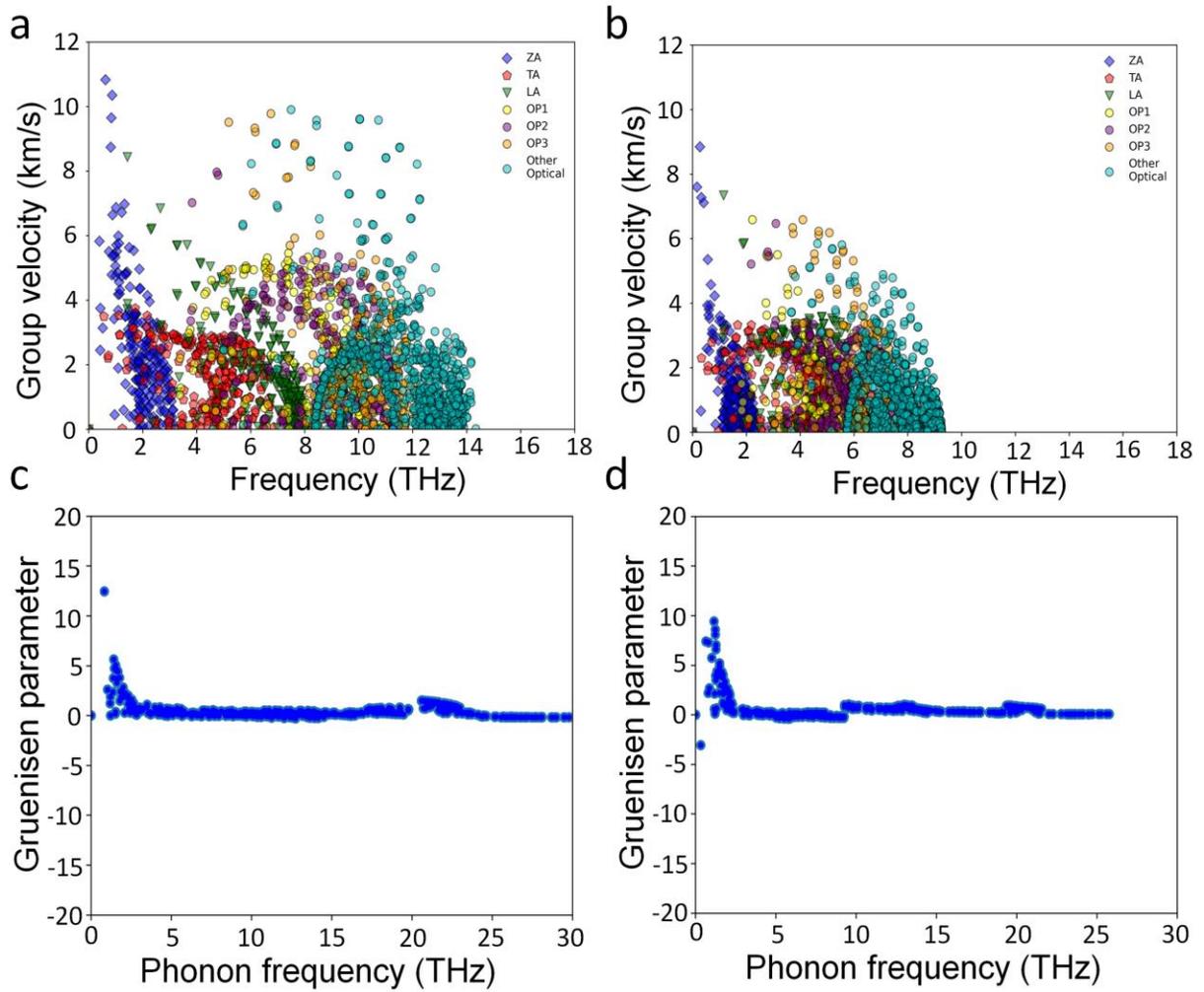

**Figure 8:** Representation of $G_v$ for different modes with respect to the frequency for (a) HfSi$_2$N$_4$ and (b) HfGe$_2$N$_4$ ML. Variation of Grüneisen parameters ($\gamma$) with phonon frequency for (c) HfSi$_2$N$_4$ and (d) HfGe$_2$N$_4$ MLs

### 3.7. Figure of merit (*ZT*)

A high *ZT* value requires a high value of *S* and $\sigma$ and a considerably lower value of $k$ ($k_e + k_l$), as indicated in the equation 8. The figure of merit (*ZT*) is defined as

$$ZT = \frac{S^2 \sigma T}{\kappa_e + \kappa_l}, \tag{8}$$

where *S* is the Seebeck coefficient, *T* is the temperature, $\sigma$ is the electrical conductivity, and $\kappa_e$ and $\kappa_l$ are the components of thermal conductivity, as stated earlier. **Figure 9 (a) and (b)** show the relation between the *ZT* and ɸ for the ML of HfSi$_2$N$_4$ and HfGe$_2$N$_4$, respectively, at different temperatures. HfGe$_2$N$_4$ ML shows a higher *ZT* value at 900 K compared to HfSi$_2$N$_4$, i.e., 0.90 and 0.83 for p-type and n-type doping, respectively. Additionally, at 900 K, the HfSi$_2$N$_4$ ML demonstrated a significant *ZT* of 0.89 for the p-type and 0.79 for the n-type. The p-type doping exhibited greater *ZT* values in comparison to the doping with n-type for both structures. The *ZT* values obtained from the HfSi$_2$N$_4$ and HfGe$_2$N$_4$ at different temperatures are shown in **Table 3**.

**Table 3:** Obtained *ZT* at different temperatures of ML of HfSi$_2$N$_4$ and HfGe$_2$N$_4$.

| Temperature | ZT | | | |
|---|---|---|---|---|
| K | HfSi$_2$N$_4$ | | HfGe$_2$N$_4$ | |
| | p | n | p | n |
| 300 | 0.55 | 0.32 | 0.68 | 0.41 |
| 600 | 0.82 | 0.66 | 0.84 | 0.73 |
| 900 | 0.89 | 0.79 | 0.90 | 0.83 |

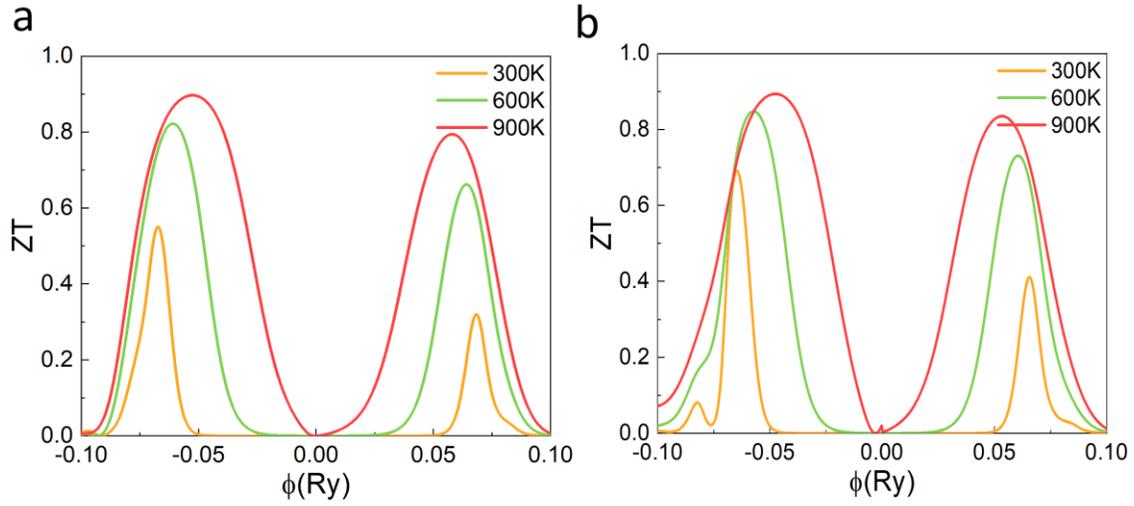

**Figure 9:** Representation of $ZT$ for different temperatures ($T$) as a function of ϕ for (a) $HfSi_2N_4$ and (b) $HfGe_2N_4$ ML.

## 4. Conclusions

We studied optoelectronic and thermoelectric properties using first-principles calculations. The structural stability for both the MLs was verified from the phonon dispersion relations which are free from imaginary frequencies. The $ZT$ for both $HfSi_2N_4$ and $HfGe_2N_4$ can be considered very good for thermoelectric applications for both low-temperature and high-temperature thermoelectric devices in the future. The $HfGe_2N_4$ showed a better $k_l$ value compared to many popular TMDCs. The $HfGe_2N_4$ showed $ZT$ values of 0.68, 0.84, and 0.90 at room temperature (300 K), 600 K, and 900 K for p-doping. So, here, p-doping is more effective than n-doping. The $S$ and $\sigma$ of $HfSi_2N_4$ is more significant rather than $HfGe_2N_4$, so the PF of $HfSi_2N_4$ is higher than $HfGe_2N_4$, But the $k_l$ of $HfGe_2N_4$ is found to be lower than $HfSi_2N_4$ so the $ZT$ of $HfGe_2N_4$ become higher. The $HfGe_2N_4$ ML can also be used as a low-temperature thermoelectric device as it showed considerable $ZT$ at room temperature.

### Acknowledgments:
We are thankful to the INSPIRE program and the Indian Institute of Technology Jodhpur for their financial assistance and necessary facilities to conduct the research.